# Pre-Seismic Electrical Signals (SES) generation and their relation to the lithospheric tidal oscillations K2, S2, M1 (T = 12hours / 14 days).

Thanassoulas, P.C., B.Sc in Physics, M.Sc – Ph.D in Applied Geophysics.

Retired from the Institute for Geology and Mineral Exploration (IGME)
Geophysical Department, Athens, Greece.
e-mail: thandin@otenet.gr – website: www.earthquakeprediction.gr

**Abstract**

It is postulated that the preseismic electric signals **(SES)** are generated by the piezoelectric mechanism applied on small rock grains - blocks during their stress load until fracturing. Specifically, the square electric train pulses are generated by the combination of a stress increase phase which generates a positive piezostimulated polarized current pulse **(PSPC)** followed, in a short time, by the stress decrease phase at fracturing level which generates a negative piezostimulated depolarized current pulse **(PSDC)**. Moreover, it is shown that the **SES** signals are closely related to the tidally triggered lithospheric stress maxima – minima. Examples of **SES** signals are presented in relation to the tidally triggered lithospheric oscillation **(k2, S2, M1)** of T = 12hours / 14 days, while some comments are made as far as it concerns their use in short-term earthquake prediction.

**1. Introduction**

The electric, seismic precursory signals which have been presented to date in the seismological literature are mainly distinguished in the following types:

a) **SES**, (Seismic Electric Signals) or "high frequency" signals.

b) **Oscillating signals**, mainly of 24 hours / 14 days period following the corresponding tidal oscillation.

c) **VLP** signals, (Very Long Period).

Examples of such signals were presented by Thanassoulas (2007) and at therein references. A thorough study of the various mechanisms which generate electrical signals, suggests that the piezoelectric is the most probable one. This is the only mechanism that justifies in total: the generation of **SES** (signals, due to higher order harmonics generated at the non-linear part of the generated potential), **Oscillating signals,** (due to the oscillating component of the stress-strain of the seismogenic area, caused, by the oscillatory effect of the Earth-tides upon the stress-strain / piezoelectric potential curve) and the **VLP** signals, (due to the total form of the generated static potential which is caused by the large scale crystal-lattice deformation).
In this particular study a quite different physical mechanism for the **SES** generation will be postulated and the correlation of the **SES** signals with the tidal oscillations **K2, S2, M1** (mainly of T = 12 hours / 14 days) will be demonstrated through specific examples, since strong EQs are not a routine event, thus prohibiting us to apply statistical validation by using a large number of statistical samples.

**2. Theoretical model of SES (Seismic, Electric Signals) generation.**

The postulated new physical model that justifies the generation of square electric train pulses integrates two well known mechanisms.
The first one is the proposed by Varotsos (2005) that concerns the generation of the piezostimulated currents. This model is shown in the following figure (2.1).

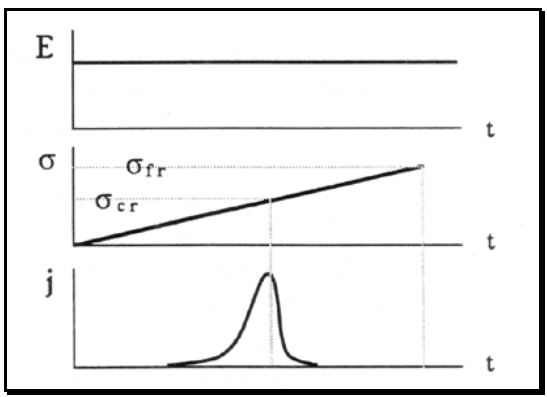

Fig. 2.1. Piezostimulated current pulse generated at a critical stress level $\sigma_{cr}$ before the fracturing $\sigma_{fr}$ one.

According to this model when a rock specimen is increasingly stress loaded, then, at a certain critical stress-level $\sigma_{cr}$, before the fracturing one $\sigma_{fr}$, generates a positive current pulse. Generally, a positive current is generated when a sudden increase of stress is applied on the rock specimen. In this case the piezostimulated current is referred as **PSPC** (piezostimulated polarizing current). The very same mechanism holds in an opposite way when the rock specimen is subjected to a sudden stress decrease. In this case, a piezostimulated depolarizing current **(PSDC)** is generated. Although this is a theoretically and practically valid mechanism, it cannot justify the generation of square train pulses which are frequently observed before large earthquakes.
The second mechanism is the piezoelectric one (Thanassoulas and Tselentis, 1986, 1993). According to this model, tidally triggered large-scale piezoelectricity in the lithosphere, produces a potential whose form is presented in the following figure (**2.2**). This potential follows closely the stress load of the seismogenic region of the lithosphere and consequently the referred seismogenic region will follow the theoretical principles of rock fracturing derived from rock mechanics.

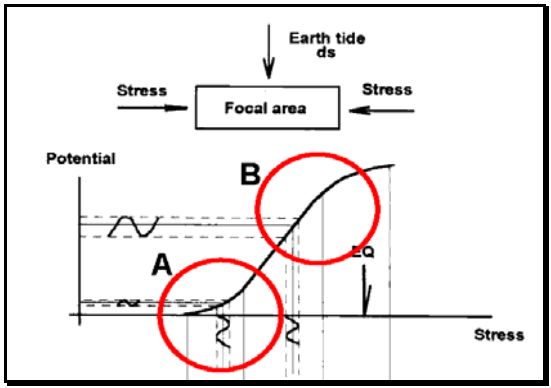

Fig. 2.2. Non-linear areas (A, B) of the piezoelectric potential curve are shown, where electric signals of higher harmonics can be produced.

The piezoelectric potential curve of figure (**2.2**) presents two regions (**A, B**) of non-linear behavior. Due to the piezoelectric mechanism these regions correspond to non-linear strain increase (for **A**) and decrease (for **B**). Therefore, in the region (**A**) since there exists a continuous non-linear strain increase, **PSPC** currents are expected to generate. In the case of (**B**) region, since there exists a continuous non-linear strain decrease, **PSDC** currents are expected to generate.

Let us now consider a very tiny rock element, which exhibits the rock properties of the seismogenic region. It is assumed that, this basic element follows the stress-strain charge conditions of the entire seismogenic area. During the boundaries of the different phases of the Mjachkin et al. (1975) model or the areas (**A**) and (**B**) of the piezoelectric model, this basic rock element undergoes, firstly, a stress-strain increase, which is followed, in very short time, by a decrease of stress-strain and fracturing, thus creating the acceleration of cracking. Therefore, this basic rock formation element is capable to produce, initially, a tiny **PSPC,** which is immediately followed by a **PSDC** before its final fracture.

Moreover, electrical signals of the same amplitude, but of different polarity, depending on the polarity (increase or decrease) of the stress-strain rate for the same basic rock element and for the same rate of stress-strain charge-discharge, will be generated.

Let us assume now that the basic rock element is so small that the time elapsed between regions **A** and **B** is small enough that a positive **PSPC** is followed immediately by a negative **PSDC**. Consequently, **each pair of positive-negative current pulses generates a "square electrical, potential pulse", the basic element of the "train pulse", which exhibits a duration that depends on the time, required, to complete its "stress load - fracture cycle".** Schematically this is presented in the following figure (**2.3**).

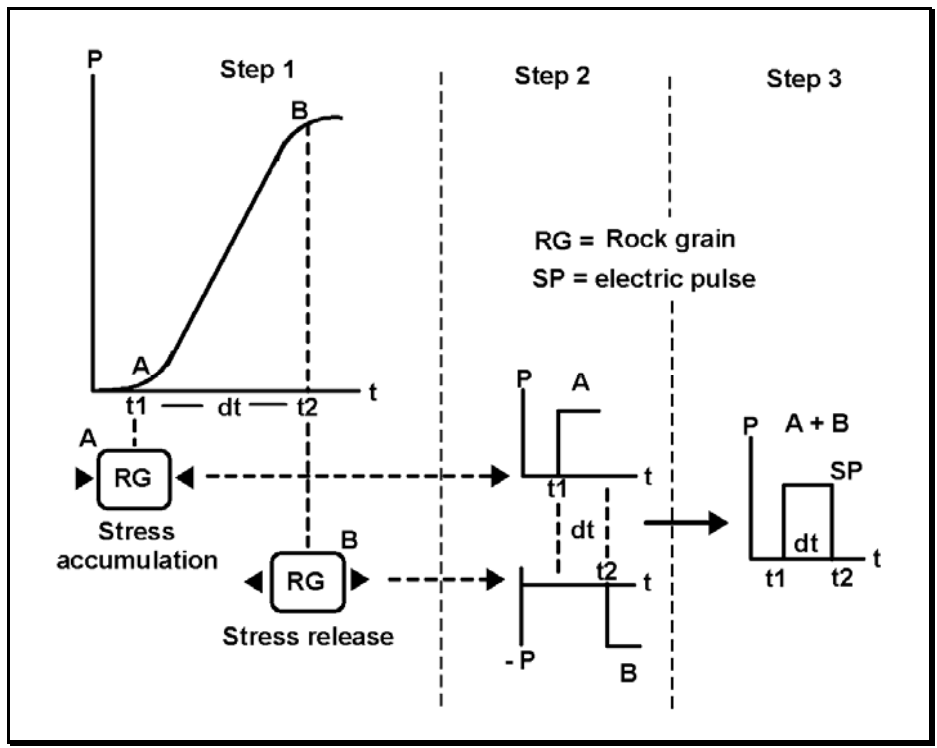

Fig. 2.3. Schematic presentation of the generation of a square electric pulse by the mechanisms of piezoelectricity and piezo-stimulated currents generation. A rock small block / grain (**RG**) is subjected in (**A**) stress increase and stress decrease in (**B**) at step (**1**). Positive (**A**) or negative (**B**) currents generate at step (**2**) which are combined in step (**3**) in a short square pulse (**SP**) of **dt** duration.

Following the later analysis, when a seismogenic area has been stress-charged at critical point, it is justified to accept that small sub-areas of it follow this stress load – fracturing cycle before its main seismic event. Therefore, electric train pulses will be generally generated. This mechanism is more intensified at lithosperic oscillation amplitude max-min times when the stress is maximized too. Consequently, at times of max – min values of the lithospheric amplitude oscillation, SES must generate, provided the seismogenic region has reaches the critical stress load point required to generate a large earthquake.

The close inspection of the **SES** signals, recorded in Greece, by **ATH, PYR** and **HIO** monitoring sites so far, indicates that a period of 8-15 minutes is typical, but shorter periods as well (from a few ms to a few minutes) have been recorded by the **VAN** group.

In the following text, the close dependence of the **SES** and the triggered by the tidal waves lithospheric oscillation will be shown.

## 3. Examples of SES vs. tidal lithospheric oscillation.

### 3.1. SES generated by the Izmit, (17/8/1999, M=7.5) EQ, in Turkey.

The electrical signals, which were recorded in Volos (**VOL**), Greece (Thanassoulas et al. 2000) before the Izmit, EQ (17[th], August, 1999, M=7.5R) in Turkey, will be the first example to be demonstrated. The detailed study of the recordings of the Earth's electrical field by (**VOL**) monitoring site, for a period of almost three months before the Izmit EQ, revealed the existence of high frequency (**SES**) electrical signals. These signals were observed almost right from the start (21[st] June, 1999) of the recording period and consequently it is justified to accept their earlier start-up time of origin.

Samples of these signals are presented bellow. During a six months period of recording of the Earth's electric field, only the **NS** component was recorded by **VOL** monitoring site. The following figure **(3.1.1)** corresponds to the recording day of 21/06/99. This recording was performed almost two months before the strong Izmit, (16/8/1999, M=7.5) earthquake in Turkey.

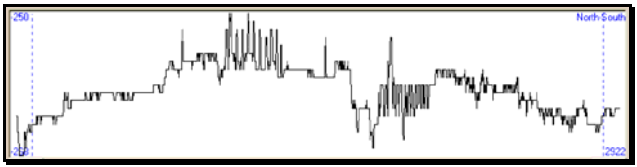

Fig. 3.1.1. Electrical signals, recorded on 21 / 06 / 1999 by **VOL** monitoring site, at **NS** direction. Note the two recorded SES.

Next data set corresponds to the recording day of 22/07/99. That is almost a month after the previous one and almost one month before the Izmit EQ.

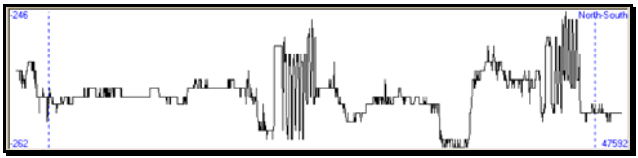

Fig. 3.1.2. Electrical signals recorded, on 22 / 07 / 1999 by **VOL** monitoring site.

Similar electrical signals have been observed in this recording, too. Almost in the entire time of recording, are observed two distinct signals, separated, at a certain period of time, of some hours. Next one was recorded on 03/08/99.

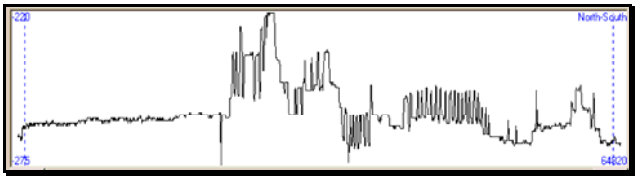

Fig. 3.1.3. Electrical signals recorded, on 03 / 08 / 1999 by **VOL** monitoring site.

The next one is a two days recording from 13 to 14/08/99. That is at a very short time, before Izmit EQ (17-08-1999) occurrence.

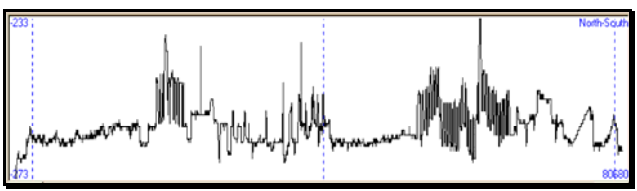

Fig. 3.1.4. Electrical signals recorded, on 13 – 14 / 08 / 1999 by **VOL** monitoring site.

The last one **(fig. 3.1.5)** had been recorded during 17-18/12/99. By that time, the most of the stress load of the seismogenic area had been released and, consequently, the **SES** signals disappeared. This is demonstrated, clearly, by the following figure **(3.1.5)**.

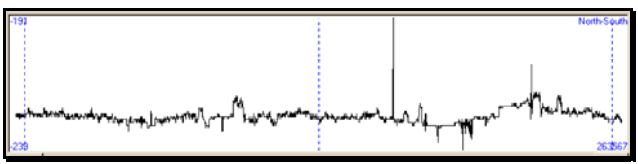

Fig. 3.1.5. Electrical field recorded, between 17-18 / 12 / 1999, by **VOL** monitoring site.

The absence of the electrical signals, comparing to the previous figures (3.1.1, 2, 3, 4), is characteristic.
Another interesting observation is that, the start time of these signals coincides with two specific daytimes.
The first one is around 9a.m, while the second one is around 21.5 p.m. This observation was studied, in detail, by separating the "9.0a.m" signals and the "21.5p.m" signals in two groups (signal – **A**, signal - **B**).
The following figure **(3.1.6)** represents the existence of electrical signals (signal – **A**) as a function of time (in days), while the vertical axis represents the startup time of each signal (in minutes) in the span of the day of its occurrence.
In the horizontal axis of time, the earthquakes of **Izmit, Athens** and **Duzce** are marked with a red arrow.

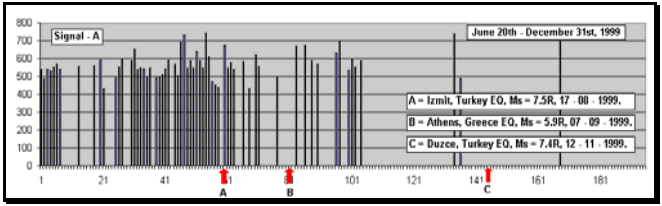

Fig. 3.1.6. Daily presence of signals – **A** is shown, during 20/06 – 31/12/1999.

In the following figure **(3.1.7),** the signals - **B** are presented with the same annotation.

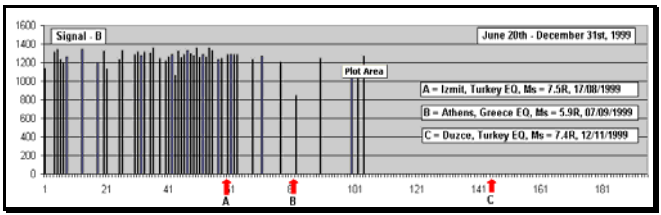

Fig. 3.1.7. Daily presence of signals – **B** is shown, for the period 20/06 – 31/12/1999.

What is clear, from both figures, is the drastic decrease of the presence of the signals after the occurrence of Izmit EQ. On the other hand, before Duzce EQ, of a similar magnitude to Izmit EQ, no such signals were observed. This suggests that Izmit – Duzce regions may be considered as a unit area, stress loaded and seismically activated. Consequently, the generated, electrical signals were produced by the entire, seismically active area and not only by Izmit focal region. This is corroborated from the fact that IZMIT – DUZCE distance is of the order of 80Km which coincides quite well with the expected fracture length of the seismogenic fault, which is what is more or less expected for an EQ of **M = 7.5**.

Therefore, when most of the stress-load of the entire area had been released (by Izmit EQ), the rest of it was not capable of generating similar, electrical signals. Viewing this pair of strong EQs from the point of view of electrical signals generation mechanism, it is a very interesting and spectacular, seismic event.

For both signals **(A, B)** the mean starting time has been calculated. For signals **(A)** the mean value **(MV$_A$)** was calculated as **569** minutes.

$$MV_A = 569 \text{ minutes} \tag{1}$$

This corresponds to a mean starting time of 9hr 29 minutes. For signals **(B),** the same calculation results in a **(MV$_B$)** of **1257** minutes. This corresponds to a mean starting time of 20hr 57 minutes.

$$MV_B = 1257 \text{ minutes} \tag{2}$$

Finally, the mean time difference in time of occurrence of the electrical signals has been calculated, as:

$$MV_B - MV_A = 11\text{hr } 28 \text{ minutes.} \tag{3}$$

Comparing this result to the Earth-tide components, its very close resemblance is revealed to **K2 (T=11.97hours, lunisolar)** and **S2 (T=12hours, principal solar)** components. A discrepancy of 4.17% has been calculated for the **(K2)** component, while a value of 4.4% corresponds to the **(S2)** one. The satisfactory results fit what was expected from the earlier theoretical analysis.

In simple words, the entire Izmit-Duzce seismogenic area was at such critically of stress-strain charged conditions point, so that it generated SES twice in a day, at the peaks of the 24 hours lithospheric oscillation.

**3.2. SES triggered by the lithospheric oscillation M1 (T = 14 days)**

Following are **(SES)** sample signals, recorded by **(HIO), (PYR)** and **(ATH)** monitoring sites, compared to the **M1**, 14 days period tidal, lithospheric oscillation local peak value.

All the **SES** signals that are presented are superimposed on top of the lithospheric tidal oscillation **M1** with **T = 14** days. A close inspection of the following figures reveals the conformity of the SES time of generation to the peaks of the **M1** lithospheric tidal oscillation.

The presented samples are grouped as recorded from **HIO, PYR** and **ATH** monitoring sites.

**3.2.1. SES signals recorded at HIO monitoring site**, compared to **M1,** 14 days period tidal lithospheric oscillation.

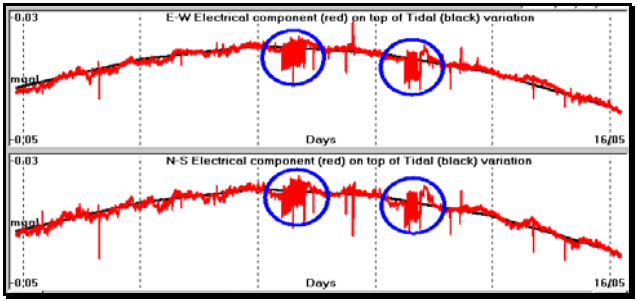

Fig. 3.2.1.1. **SES** precursory, electrical signal (in blue circles) recorded, by **HIO** monitoring site, between 13 and 14 May 2006 (HIO060513 -14).

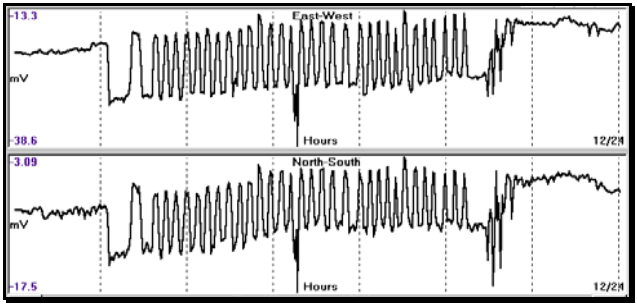

Fig. 3.2.1.1.a. Zoom-in of the left **SES** signal of fig. **(3.2.1.1)**.

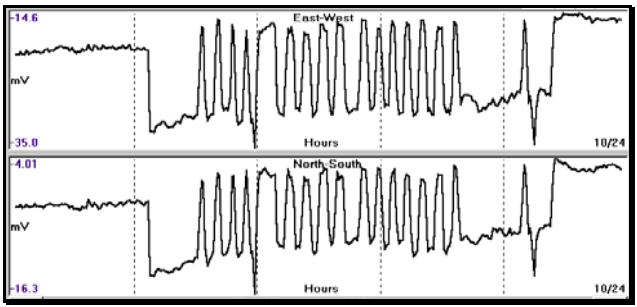

Fig. 3.2.1.1.b. Zoom-in of the right **SES** signal of fig. **(3.2.1.1).**

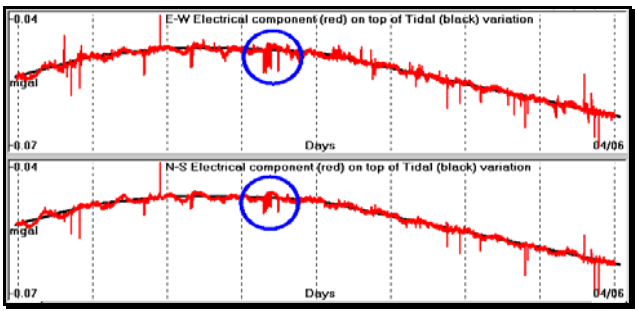

Fig. 3.2.1.2. **SES** precursory, electrical signal (in blue circles) recorded, by **HIO** monitoring site, during the 30th May 2006 (HIO060530).

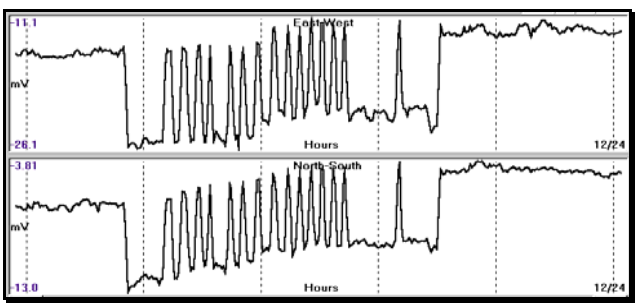

Fig. 3.2.1.2.a. Zoom-in of the SES signal of fig. **(3.2.1.2)**.

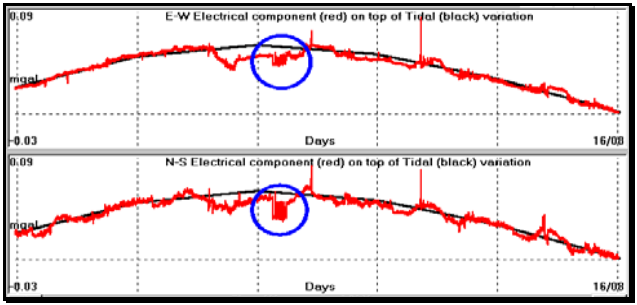

Fig. 3.2.1.3. **SES** precursory, electrical signal recorded, by **HIO** monitoring site on 13$^{th}$ August 2006 (HIO060813).

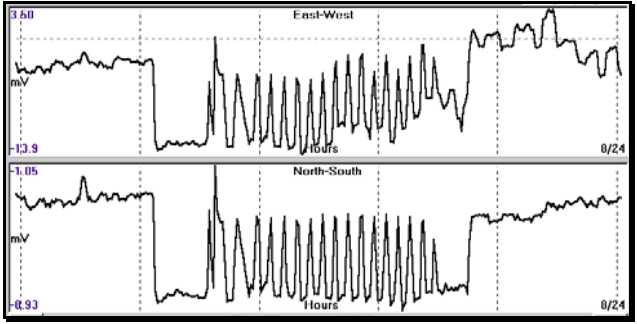

Fig. 3.2.1.3.a. Zoom-in of the **SES** signal of fig. **(3.2.1.3)**.

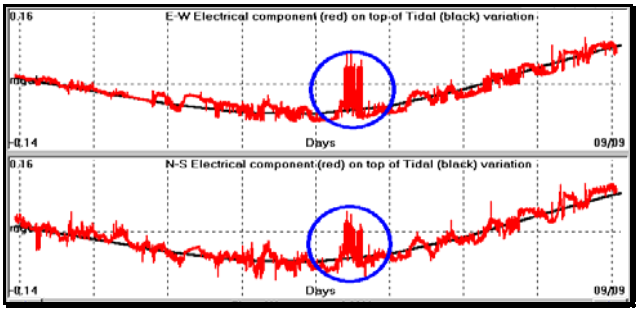

Fig. 3.2.1.4. **SES** precursory electrical signal (in blue circles) recorded by **HIO** monitoring site, on 5th September 2006 (HIO060905).

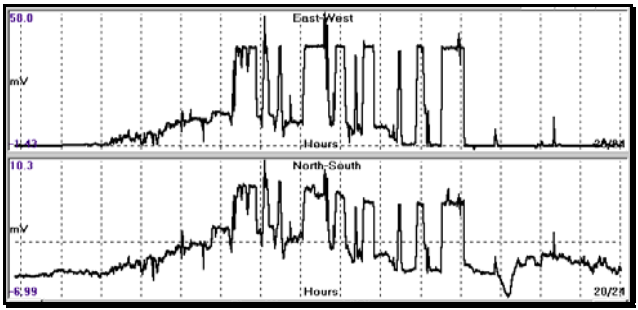

Fig. 3.2.1.4.a. Zoom-in of the **SES** signal of fig. **(3.2.1.4)**.

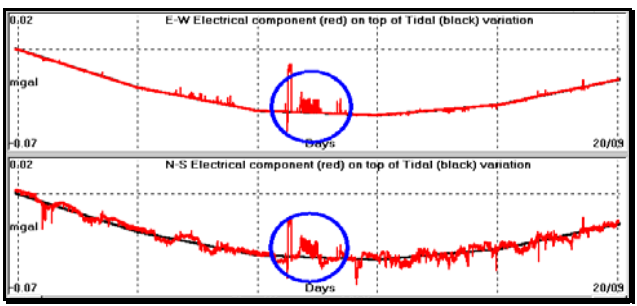

Fig. 3.2.1.5. **SES** precursory, electrical signal (in blue circles) recorded, by **HIO** monitoring site on 17th September 2006 (HIO060917).

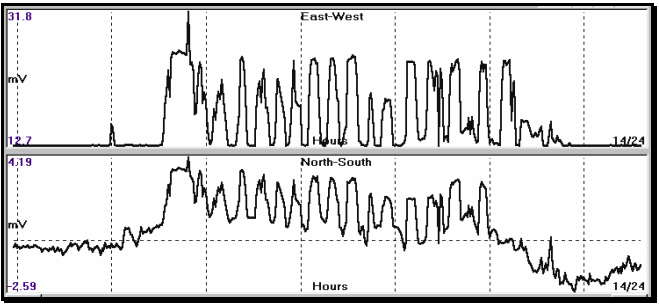

Fig. 3.2.1.5.a. Zoom-in of the **SES** signal of fig. **(3.2.1.5)**.

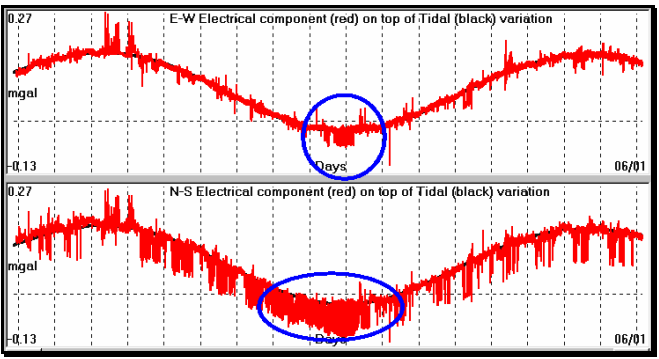

Fig. 3.2.1.6. **SES** precursory, electrical signal (in blue circles) recorded, by **HIO** monitoring site, on 28th December 2006 (HIO061228).

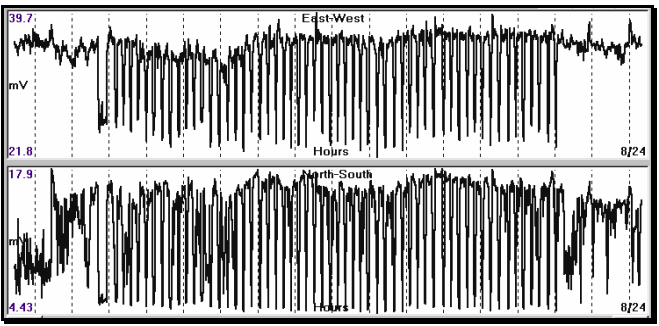

Fig. 3.2.1.6.a. Zoom-in of the **SES** signal of fig. **(3.2.1.6)**.

**3.2.2. SES signals recorded at PYR monitoring site,** compared to **M1**, 14 days period, tidal lithospheric oscillation.

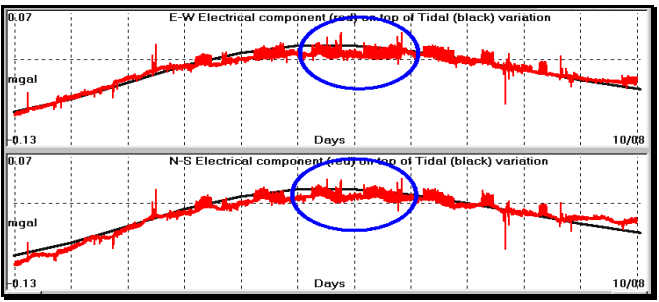

Fig. 3.2.2.1. **SES** precursory, electrical signal recorded, by **PYR** monitoring site, on 4th August 2004 (PYR040804).

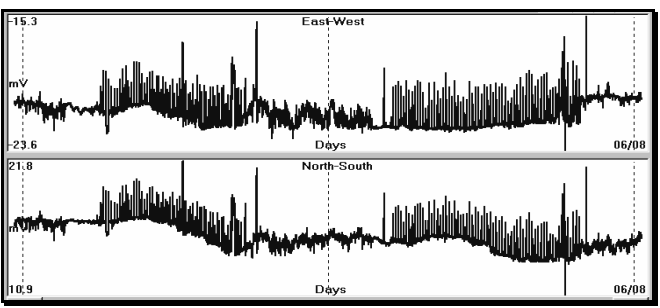

Fig. 3.2.2.1.a. Zoom-in of the (4$^{th}$ - 5$^{th}$ of August) **SES** signal of figure (**3.2.2.1**).

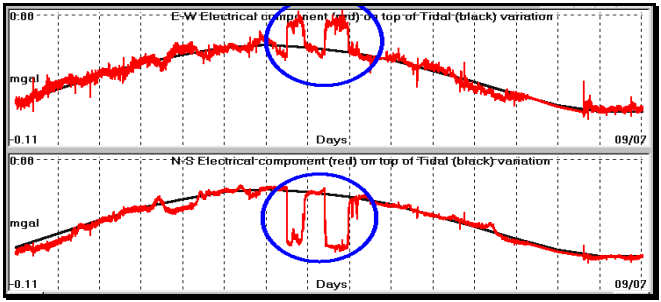

Fig. 3.2.2.2. **SES** precursory, electrical signal recorded, by **PYR** monitoring site, on 30th June, 2006 (PYR060630).

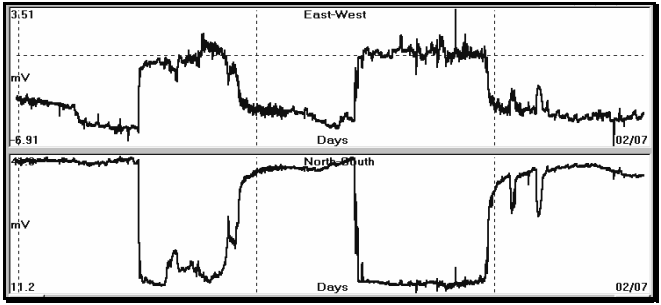

Fig. 3.2.2.2.a. Zoom-in of the **SES** signal of figure **(3.2.2.2)**.

**3.2.3. SES signals, recorded, by ATH monitoring site,** compared to **M1**, 14 days period tidal lithospheric oscillation.

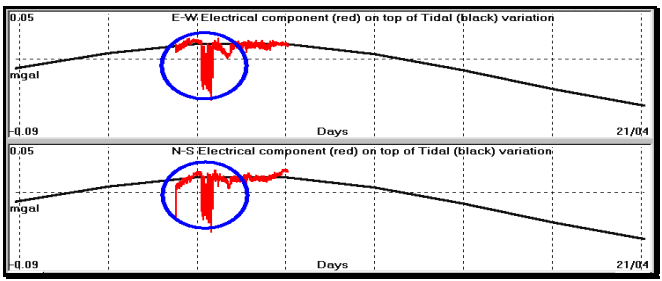

Fig. 3.2.3.1. **SES** precursory, electrical signal recorded, by **ATH** monitoring site, on 16th April, 2003 (ATH030416).

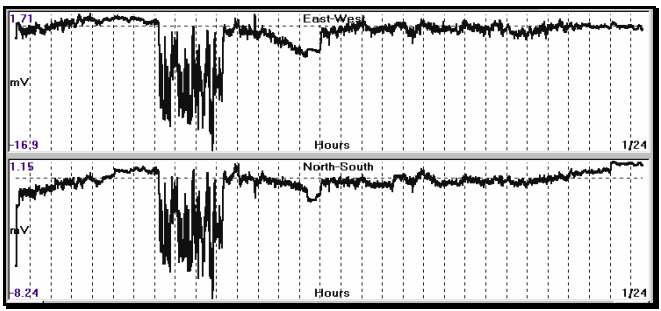

Fig. 3.2.3.1.a. Zoom-in of the SES signal of figure **(3.2.3.1)**.

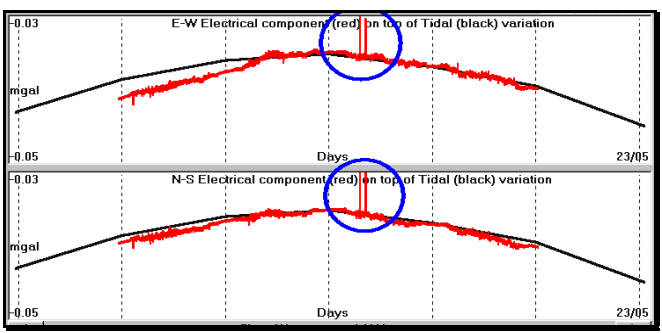

Fig. 3.2.3.2. **SES** precursory, electrical signal recorded, by **ATH** monitoring site, on 20th May, 2004 (ATH040520).

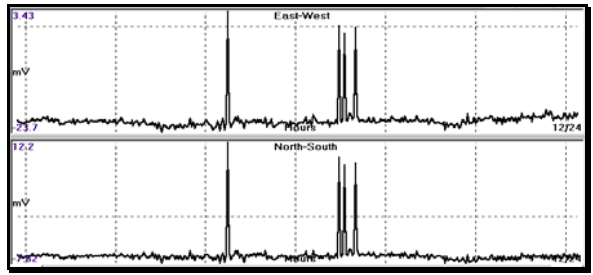

Fig. 3.2.3.2.a. Zoom-in of the SES signal of figure **(3.2.3.2)**.

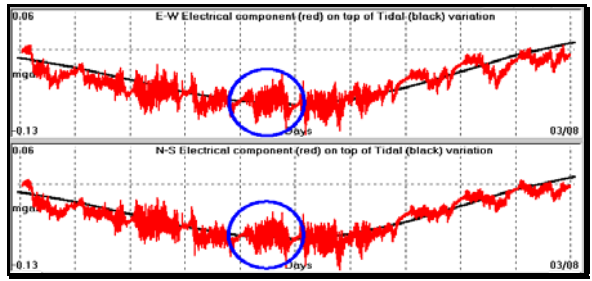

Fig. 3.2.3.3. **SES** precursory, electrical signal recorded, by **ATH** monitoring site, on 29th July, 2004 (ATH040729).

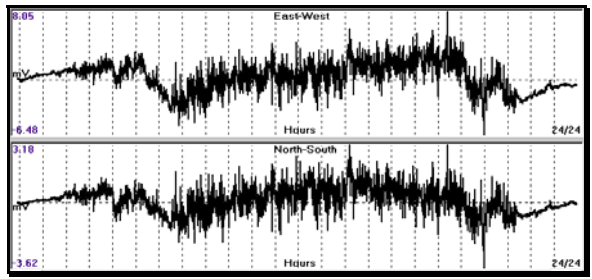

Fig. 3.2.3.3.a. Zoom-in of the SES signal of figure **(3.2.3.3)**.

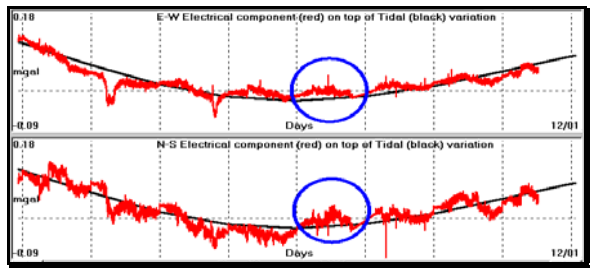

Fig. 3.2.3.4. **SES** precursory, electrical signal recorded, by **ATH** monitoring site, on 8th January, 2006 (ATH060108 EQ6.9).

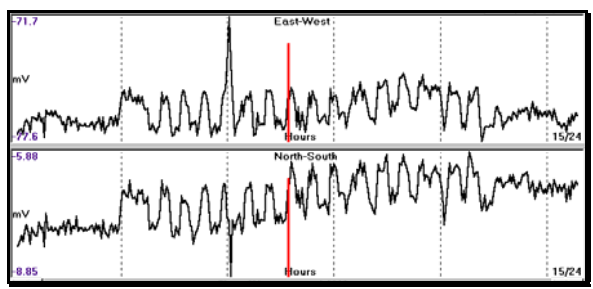

Fig. 3.2.3.4.a. **SES** precursory, electrical signal recorded, by **ATH** monitoring site, on 8th January, 2006 (ATH060108 EQ6.9). The red bar indicates the time of occurrence of the 6.9R EQ (ATH060108 EQ6.9).

The previous figures **(3.2.3.4 – 3.2.3.4.a)** indicated the close correlation of the EQ time of occurrence, to the **M1,** 14days tidal, lithospheric oscillation peak time and the present "coseismic" **SES** signal.

It developed almost 90 minutes before the EQ occurrence and vanished, almost 110 minutes after it. This suggests that, the catastrophic deformation of the seismogenic area is estimated to have lasted for about 3 hours.

### 3.2.4. SES recorded by the VAN network.

In the literature, and specifically in the papers which were published by the **VAN** group during their research activity, a lot of **SES** signals have been presented. It is worth to test the time (day) of occurrence of these signals against the **M1** tidal lithospheric oscillation of the 14 days period. To this end, the already published signals which were traceable (**No = 64**), are tabulated in **Table - 1**. The first column indicates the time of occurrence of the **SES** in yyyymmdd format, while the second one indicates the time lag, between the **SES** times considered as "zero" time and the time of the tidal peak value.

**TABLE - 1**

| SES date (yyyymmdd) | Tidal peak lag (days) | SES date (yyyymmdd) | Tidal peak lag (days) |
|---|---|---|---|
| 19810707 | -2 | 19810923 | 0 |
| 19810924 | 0 | 19810927 | -4 |
| 19810929 | 2 | 19810930 | 1 |
| 19811001 | 0 | 19811219 | 0 |
| 19820118 | 0 | 19830115 | 0 |
| 19830118 | -2 | 19830130 | 0 |
| 19830131 | -1 | 19830202 | -3 |
| 19830217 | -3 | 19830607 | -1 |
| 19830614 | -1 | 19830704 | 2 |
| 19830711 | 1 | 19831007 | 1 |
| 19831010 | -1 | 19840123 | 2 |
| 19840207 | -3 | 19840208 | -2 |
| 19840219 | -1 | 19840504 | -3 |
| 19840508 | 0 | 19840514 | 0 |
| 19840515 | 0 | 19841114 | -2 |
| 19850412 | 0 | 19850430 | -1 |
| 19850514 | -2 | 19850516 | 3 |
| 19880401 | 0 | 19880515 | 1 |
| 19880831 | -1 | 19880831 | 0 |
| 19880929 | -1 | 19881003 | 2 |
| 19891018 | -1 | 19900426 | -1 |
| 19930127 | -3 | 19930224 | -2 |
| 19930224 | -2 | 19950406 | 2 |
| 19950418 | -1 | 19950418 | -1 |
| 19950419 | 3 | 19950430 | 0 |
| 19950513 | 2 | 19950917 | 2 |
| 19980621 | -1 | 19980814 | -2 |
| 19990901 | -1 | 19990902 | -2 |
| 19990906 | 0 | 19990907 | 0 |
| 20010317 | 0 | 20010725 | -1 |
| 20040514 | -2 | 20050321 | 3 |
| 20050407 | 0 | 20060323 | 0 |

The negative values, of the second column, indicate that the **SES** time of occurrence follows the **M1** tidal peak", while the positive values indicate that the SES "preceded the corresponding **M1** tidal peak".

The values, tabulated, in table **(1)**, are presented in a graph form, in the following figure **(3.2.4.1)**.

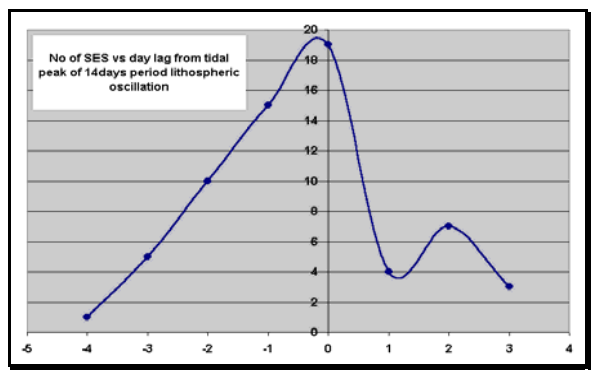

Fig. 3.2.4.1. Number of **SES** are presented as a function of time lag to tidal peak of the 14 days lithospheric oscillation.

The form of the function of **SES** time vs. time lag to **M1** tidal oscillation peak times suggests that the majority of the **SES** occur at some time "following" the **M1** tidal peak occurrence. By taking into account that the stress-strain charge of the lithosphere is at maximum load conditions during peak tidal oscillating values, then it is justified to accept that the majority of **SES** are generated in "decompression conditions" or in other words **SES** are **"pressure stimulated depolarization currents - PSDC".**

The stress-strain charge conditions where an **SES** may develop are regions of increase or decrease of the stress-strain load of the lithosphere. It is worth to compare such stress-strain changes in the lithosphere, during the preparation phases of an earthquake. Firstly, is considered the model, proposed, by Mjachkin et al. (1975). In this model **(fig. 3.2.4.2)**, three characteristic phases are present, concerning the deformation velocity during a seismic cycle. In the first **(I)** phase, the seismogenic area exhibits homogeneous cracking; during the second **(II)** phase, cracking acceleration, due to interaction of cracks, is exhibited; and during the third **(III)** phase, unstable cracking and main fault formation is exhibited, and the generation of the main, seismic event follows.

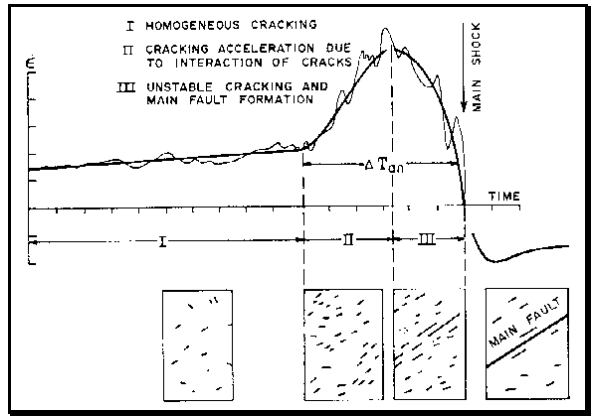

Fig. 3.2.4.2. Change of average deformation velocity, during the seismic cycle (Mjachkin et al. 1975).

Change of stress-strain load, which can cause the generation of **SES,** occurs in the boundaries between phases **(I)** and two **(II)**, between phases **(II)** and **(III)** and very shortly before the main, seismic event. **The time span of each phase is generally unknown and therefore, even if the SES has been observed, the remaining time window for the occurrence of the pending seismic event, is still highly unpredictable.**

The piezoelectric model will be considered next (fig. **3.2.4.3**). There are two distinct regions, where non-linear change of the strain load exists in the strain-stress curve. In area **(A),** there is a rapid non-linear increase of the strain, while in area **(B),** there is a non-linear decrease of the strain. These are favorable areas, where **SES** can develop either as **PSPC at (A)** or as **PSDC at (B)**.

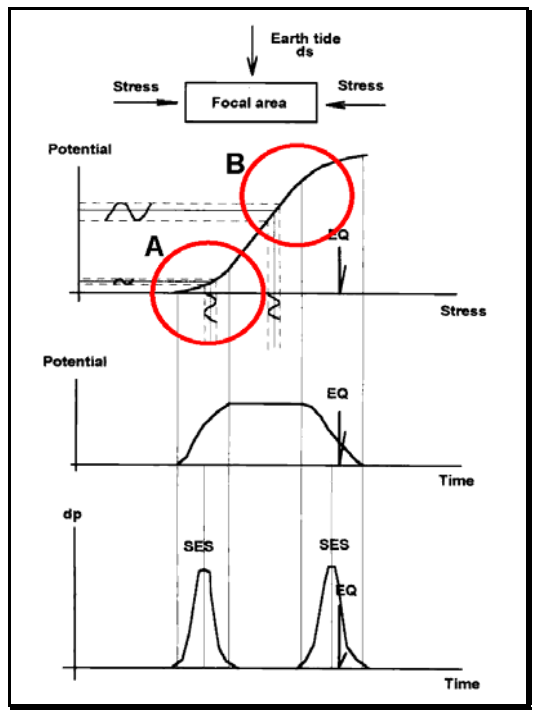

Fig. 3.2.4.3. Areas of "compressional **(A)** stress - strain increase" where **PSPC** can generate and "decompressional **(B)** stress - strain decrease" where **PSDC** can generate.

These signals can be considered as higher order harmonics of the total piezoelectric potential generated, in terms of the function of Piezoelectric Potential vs. time / stress load. **The timing problem of an earthquake still exists, since the time interval between areas (A) and (B), is still unknown.**

A search, in the **VAN** group literature, indicates that the time of occurrence of an **SES** may precede the time of occurrence of the corresponding earthquake, from **some minutes** (papers of 1981) up to **11 days** (papers of 1993) and probably of longer periods of time. The later is not surprising, if it is explained by the tidal lithospheric oscillation. Actually, an **SES** can be generated at any favorable tidal oscillation peak value, but the corresponding earthquake will occur later on, when its critical stress-strain conditions are met at a specific, in future, tidal oscillation peak.

## 4. Conclusions.

Summarizing all the above, it can be said that:

Preseismic electric signals (**SES**), in the form of train-pulses, can be produced by the combined effect, on small rock blocks, of piezostimulated currents generation mechanism and the piezoelectric one in a larger scale.

The **SES** presence is a clear indication that an unknown seismogenic area has reached a stress-strain level, close, to the critical stress-strain charge conditions, which are necessary for the generation of a large seismic event.

The **SES** signals do generate, at their vast majority, at the peaks of the lithospheric oscillation triggered by the tidal waves. At their majority these originate as **PSDC** preseismic electric signals.

The presented examples suggest that the **SES**, when the seismogenic region is at a critical stress load point, can be triggered by the **K2, S2** and **M1** tidal components.

However, the problem that still exists, in terms of "prediction", is that, **there is no method, based solely on the SES presence (in terms of short-term time prediction), to suggest the deterministic calculation of the remaining time between the SES occurrence and the time of occurrence of the future seismic event.**

**When this seismic event will occur, is still unknown. Actually it takes the study of some more physical properties and the use of combined mechanisms (Thanassoulas, 2007) which will shorten the suggested time window for a truly short-term earthquake prediction.**

## 5. REFERENCES.